\documentclass[article]{elsarticle}

\newcommand{\mathbb}[1]{\mathbf{#1}}

\usepackage{media9}
\usepackage{multimedia}
\usepackage{media9}

\RequirePackage{color}
\definecolor{MyDarkGreen}{rgb}{0.02,0.60,0.06}

\begin{document}

\title{Spreading processes in ``post-epidemic'' environments. II.  {Safety    patterns} on scale-free networks}

\author[a,b]{V. Blavatska\corref{cor}}\ead{blavatskav@gmail.com}
\author[a,b,c]{Yu. Holovatch}
\cortext[cor]{Corresponding author}
\address[a]{Institute for Condensed Matter Physics of National Acad. Sci. of Ukraine, Lviv, Ukraine}
\address[b]{${\mathbb {L}}^4$  Collaboration \& Doctoral College for the Statistical Physics of Complex
Systems, \\Leipzig-Lorraine-Lviv-Coventry, Europe}
\address[c]{Centre for Fluid and Complex Systems, Coventry University,\\ Coventry, CV1 5FB,
United Kingdom}

\date{\today}

\begin{abstract}
This paper continues our previous study on spreading processes 
in inhomogeneous populations consisting of susceptible and immune individuals
[V. Blavatska, Yu. Holovatch, Physica A {\bf 573}, 125980 (2021)].
A special role  in such
populations is played by ``safety patterns'' of susceptible nodes surrounded by 
the immune ones.
Here, we analyze spreading on  scale-free networks, where the distribution of node
connectivity $k$ obeys a power-law decay $\sim k^{-\lambda}$. We assume, that only 
a fraction $p$ of individual nodes can be
affected by spreading process, while remaining $1-p$ are immune.  We apply the synchronous cellular 
automaton algorithm and study the stationary states and spatial patterning in SI, SIS 
and SIR models in a
range  $2 < \lambda < 3 $.  Two immunization scenarios, the
random immunization and an intentional one, that targets the highest degrees nodes  
are considered. 
A distribution of safety patterns 
is obtained for the case of both scenarios. Estimates for the threshold values of the 
effective spreading rate $\beta_c$ as a function of
active agents fraction $p$ and parameter $\lambda$ are obtained and efficiency of both 
vaccination techniques are
analyzed quantitatively.  The impact of the underlying network heterogeneous structure  is manifest e.g. in decreasing the  $\beta_c$ values within the random scenario 
as compared to corresponding values in the case of regular latticek. This result quantitatively 
confirms the compliency of  scale-free networks for disease spreading. On contrary, the 
vaccination within the targeted scenario makes the complex networks much more resistant to
epidemic spreading as compared with regular lattice structures.

\end{abstract}


\maketitle

\section{Introduction} \label{I}

The concept  of complex networks is widely used to describe
both the  nature and society; many systems can
be treated as  graphs with nodes representing individuals
or agents and links displaying  any kind of mutual interactions between them. 
The question of great interest  it thus the impact of complex network topology  on
the dynamics of  spreading phenomena on such structures
\cite{Moore01}, ranging from computer
virus \cite{Murray88,Newmann02} to rumour spread \cite{Daley65} and to  {infectious} diseases \cite{Murray93}.
Of particular importance in this respect are the so-called scale-free (SF) networks 
\cite{Barabasi99},  where the node degree distribution $P(k)$ is governed by  a power-law decaying:
\begin{equation}
 {P(k)}\sim  k^{-\lambda}, \hspace{2em} k\gg1\, . \label{power}
\end{equation}
It is established, that a number of important real networks such as  WWW  \cite{Faloutsos99,Caldarelli00, Goh02},  
collaboration and co-authorship \cite {Barabasi99, Newman01},  food web and trophic interactions in ecosystems \cite{Navia16}  etc. 
are characterized by distribution (\ref{power}) with exponent $\lambda$ in a range $2 < \lambda < 3$.

The spreading processes on SF networks have been thoroughly analyzed, in particular,  within  the frames of 
archetype  scenarios of epidemics invasion (based on the so-called compartment models \cite{Kermack27,Anderson92} considered in more details in 
Section \ref{Spreadingtypes}).
Such  networks are found to be highly compliant to 
spreading  processes \cite{Pastor01a,Pastor01b,Pastor02,Pastor15}. It is caused by
diverging connectivity fluctuations of SF networks with $\lambda < 3$ (in the asymptotic limit of infinitely large  networks,  
the second moment $\langle k^2 \rangle$ of the degree distribution (\ref{power}) diverges) and by
a statistically significant probability for some nodes to have a very high degree compared to the mean one.
Note however that for smaller size networks  $\langle k^2 \rangle$  has  a large but finite value, defining an effective 
nonzero threshold for epidemic outbreak in such systems due to the finite
size effects, as it is usual in  non-equilibrium phase
transitions \cite{Marro99}. 
An impressive research effort has been devoted to deeper
understanding of the spreading dynamics  on complex SF networks, 
including degree-based mean-field theories \cite{Pastor01a,Pastor01b,Boguna02,Morreno02,Morreno03,Bart05}, exact methods 
\cite{Mieghem13,Chatterjee09} and numerical simulations \cite{Morreno02,Morreno03,Ferreira12,Mata13}, see Ref. \cite{Pastor15} for a recent review.

 Since the infection with $100\%$  efficiency is
not realistic even for highly
virulent diseases, it is reasonable to use a model
in which only some fraction $p$ of nodes
are considered susceptible to the disease transmission. One can observe such a case   e.g. 
in  population, where $1-p$ individuals  has become immune, whether through vaccination or being infected and cured  previously.
Immunization strategies  protect the population from
the global propagation of a disease and can also lead
 to an increase of the
epidemic threshold  (this effect is called herd immunity) \cite{Fine11}.	
In our previous work \cite{Blavatska21}, such a situation was considered
 by studying the stationary states and spatial patterning  on
	a square lattice with the fraction $p$ of  {susceptible to disease transmission (active)} sites. An emergence of  
	``safety patterns'' of susceptible agents surrounded by immune individuals
	in such a system was described quantitatively. This concept plays an important role in the course of epidemic processes 
	and determines the fraction of infected agents in a stationary state.
	Estimates for the threshold values of epidemic outbreak as a function of active agents fraction $p$ on a square lattice 
	were obtained by us as well. 
	
The present work serves as a continuation of the study,  {initiated in \cite{Blavatska21} generalizing it for the case} of SF 
networks with node degree distribution given by (\ref{power}) 
  with exponent $\lambda$ in a range $2 < \lambda < 3$. 
The fraction $p$
of nodes are considered as susceptible for  disease spreading, whereas
the rest $1-p$ are treated as immune. 
Various  immunization scenarios can be performed in such networks due to their heterogeneity. The simplest  (and not effective one)
is the random immunization, in which a number $(1-p)N$ of nodes is randomly 
chosen and made immune.  For the infinite networks it was proven \cite{Pastor02b} that  
almost the whole network must be immunized to suppress
the disease.
  More effective level of protection in SF networks is achieved by means of optimized
immunization strategies, targeting the  highest degree nodes  \cite{Cohen01,Pastor02b,Cohen03},
which  are potentially the strongest spreaders.
Another
effective targeted immunization strategies are based e.g. on the
betweenness centrality, which combines the ideas of taking into account the  highest degree nodes and  the most
probable paths for infection transmission \cite{Holme02}. The  immunization protocols  can be improved
also by  allowing for each node
to have  {information about the} degree of its nearest
neighbors, and thus immunizing the
neighboring nodes with the largest degree \cite{Holme04}.

The layout of the paper is as follows.  In the next Section \ref{II}, we introduce the model for constructing the SF networks with desired power-law 
distribution (\ref{power}) and analyze the general impact of vaccination (removing 
some amount of the active nodes) on network connectivity. In Section \ref{III},  we briefly overview possible spreading scenarios and 
describe the cellular automaton  method, which is applied to implement them. Our main results for 
the quantitative description of disease spreading within two different vaccination scenarios are given in Section \ref{IV}. 
We end up with giving Conclusions in Section \ref{V}.


\section{The Model} \label{II}

Aiming to study the infection spreading process on SF networks with a power-law degree distribution in the form (\ref{power}), 
we make  use of the so-called configuration model to construct the network and then 
apply two different scenarios to 
 make some amount of nodes  considered as vaccinated.

\subsection{Constructing a network}

Within the frames of the configuration
model  \cite{Bender78,Molloy95}, we start with a set of  $N=6000$ disconnected nodes.   
A degree $k_i$ ($i=1,\ldots,N$) is assigned to each of the nodes ($\sum_i k_i$ should be an even number, since each link should connect the  two nodes). 
The degree is given by a random number selected from probability distribution $P(k)$ with condition $k_{{\rm min}}\leq k_i \leq k_{{\rm max}}$ 
with desired minimal and maximal 
degrees, correspondingly. We take $k_{{\rm min}}=2$  to provide connectivity of network (the existence of so-called giant connected component)  
\cite{Molloy98,Cohen00}. 

The maximum degree cutoff $k_{{\rm max}}\sim N ^{1/2}$ is introduced  \cite{Catanzaro05} in order to decrease the degree correlations 
(disassortativity) in configuration model.
The actual network is constructed by randomly connecting the nodes  according to the prescribed numbers of their outgoing links $k_i$  with control 
of avoidance of multiple connections and self-connections.  The resulting network thus contains $\sum_{i}k_i/2$ links. 

\subsection{Network gets vaccinated}\label{rozpodilSec}

As the next step, we assume that some fraction of nodes  is immune (these nodes together with their outgoing links  are removed from the lattice).	
In this respect, we recall the problem of the so-called node  percolation  on a network.
The concept  of the threshold  probability
$p_{{\rm perc}}$ naturally arises in such systems,  such that at the concentration of remaining nodes $p<p_{{\rm perc}}$ the network is composed of a set of isolated
disconnected clusters whereas at $p>p_{{\rm perc}}$ a giant cluster spans the entire
network \cite{Molloy95,Molloy98}. 
The percolation thresholds $p_{{\rm perc}}(\lambda)$ can be estimated  on the basis of the
general Molloy-Reed criterion for the existence of a spanning cluster \cite{Cohen00}:
\begin{equation}
\frac{\langle k^2 \rangle(p_{{\rm perc}})}{\langle k \rangle(p_{{\rm perc}})}=2, \label{criterion}
\end{equation}
where $\langle k \rangle(p_{{\rm perc}})$, $\langle k^2 \rangle(p_{{\rm perc}})$
are the mean and mean square node degree calculated at the percolation threshold.

While at $\lambda>3$ the percolation threshold takes finite values, in the case of infinitely large networks
with $\lambda\leq3$
a spanning cluster exists at any arbitrarily small value $p>p_{{\rm perc}}\approx0$.
In finite systems, however, a
percolation transition is observed at $\lambda\leq 3$ too, although the transition threshold $p_{{\rm perc}}$
is very small.


In what follows, we consider 
 two different immunization scenarios that target either the random chosen nodes or the most ``important'' nodes, as explained in more details below. 
{  Let us note also that in what follows  
the averaging of all observables of interest is performed for each fixed $p$ value  over an ensemble of $1 000$ replicas of network realizations.
}

 \subsubsection{Random scenario}

In the simplest scenario,  we chose at random a fraction of $1-p$  nodes and consider them as vaccinated.
As  a result, the network contains  a number of  subgraphs of $s$ linked susceptible nodes, considered   as clusters of size $s$. 
To extract clusters of different sizes numerically, we apply an algorithm
developed by Hoshen and Kopelman \cite{Hoshen}.   {   This algorithm is successfully applied in studies of percolation  phenomenon  in disordered environments \cite{Blavatska08, Lapshina19,Kotwica19,Oliveira20,Stauffer} . 
}

\begin{figure}[h!]
	\begin{center}
	        	\includegraphics[width=70mm]{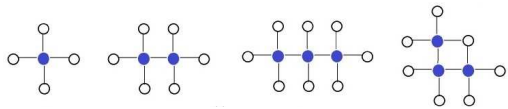}	
		{  
		\caption{(Color online) Possible configurations of clusters of susceptible nodes (dark discs,  blue online) up to the size $s=3$ on a square lattice.}
		}
\label{1}	
\end{center}
\end{figure}

\begin{figure}[h!]
	\begin{center}
	        	\includegraphics[width=70mm]{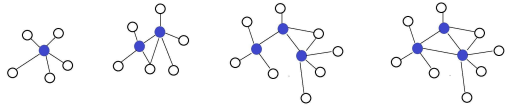}	
	{  	\caption{(Color online) Possible configurations of clusters of susceptible nodes (dark discs,  blue online) up to the size $s=3$ on a network.}}
\label{2}	
\end{center}
\end{figure}

{
For cluster containing $s$ nodes, let us introduce the number of such clusters per one node $n_s(p)$  \cite{Stauffer}
 (the probability $P_s(p)$ to find a cluster containing $s$ susceptible nodes is thus
 $P_s(p)=sn_s(p)$).
 Let us start with considering the case of simple square lattice 
with fraction $p$ of susceptible sites (see Fig. 1). 
For a single  susceptible site surrounded by the immune ones, as shown on the leftmost panel of Fig. 1,  $n_1(p)$ is given as a
probability $p$ to find a susceptible site times probabilities
of the four neighboring sites to be immune $(1-p)^4$, so that
\begin{equation}
n_1(p) = p(1-p)^4.
\end{equation}
For the  cluster of two sites, the second panel of Fig. 1, $n_2(p)$ can be calculated as a probability of two neighboring sites to be active $p^2$ and their six neighbors to be immune $(1-p)^6$  taking into account that there are two such configurations (the cluster can be oriented horizontally or vertically), so that
\begin{equation}
n_2(p) = 2p^2(1 -p)^6.
\end{equation}
For the  cluster of three sites, the third and the fouth panels of Fig. 1,  there are  two possible configurations, and number of such configurations is correspondingly 2 and 4, resulting in
\begin{equation}
n_3(p) = p^3(2(1 -p)^8+4(1 -p)^7). 
\end{equation}
Applying such considerations, exact values for
$n_s(p)$ with $s$ up to 17 were obtained in Ref. \cite{Sykes} for a simple square lattice.
For the case of complex network, the number of nearest neighbors for each node is not fixed.  Following the same reasoning as those given above in the simple square lattice case, one can estimate the number of single-node ($s=1$) clusters on a network as
\begin{equation}
n_1(p) = p(1 -p)^{\langle k \rangle}
\end{equation}
 with the mean node degree $\langle k \rangle$  obtained from the corresponding node distribution,
while for $s=2$ it can be approximately estimated as
\begin{equation}
n_2(p) = p^2(1 -p)^{2\langle k \rangle-2}.
\end{equation}
  The exact values for $n_s(p)$ with larger $s$ are tricky to calculate due to  compicated mutual interlinking within the network.
}
 \begin{figure}[t!]
	\begin{center}
	        	\includegraphics[width=70mm]{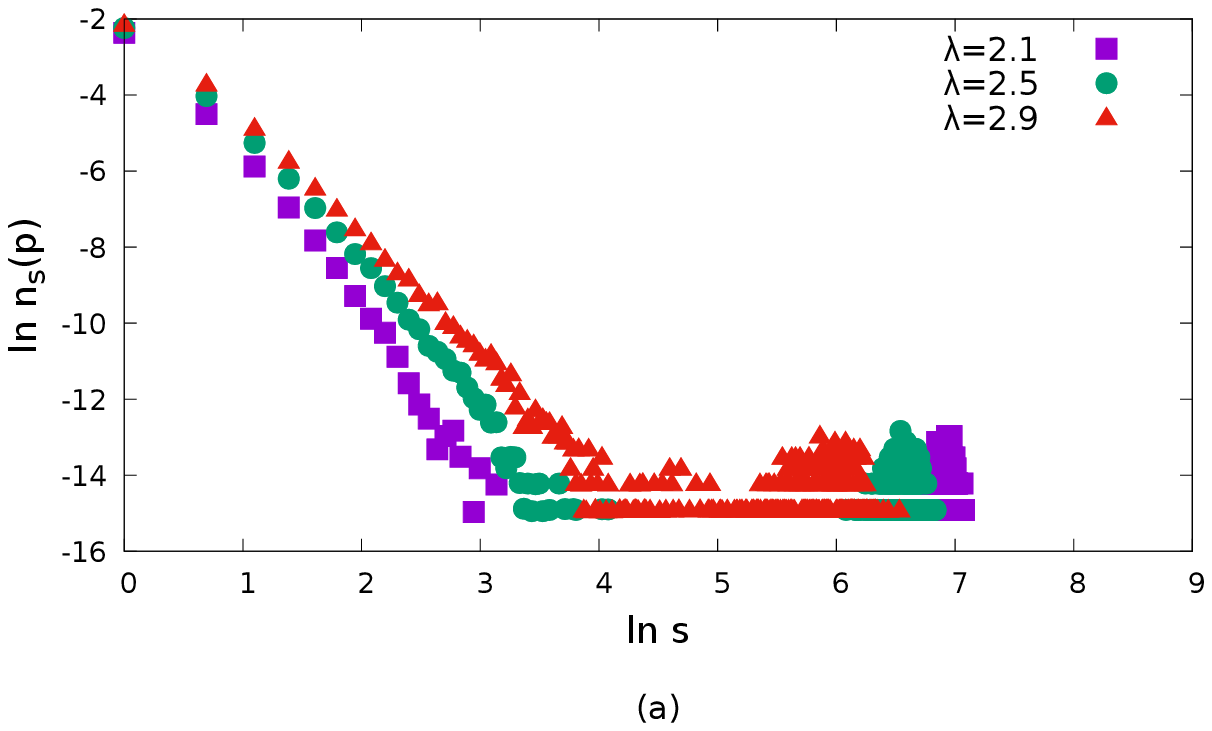}	
	 	\includegraphics[width=70mm]{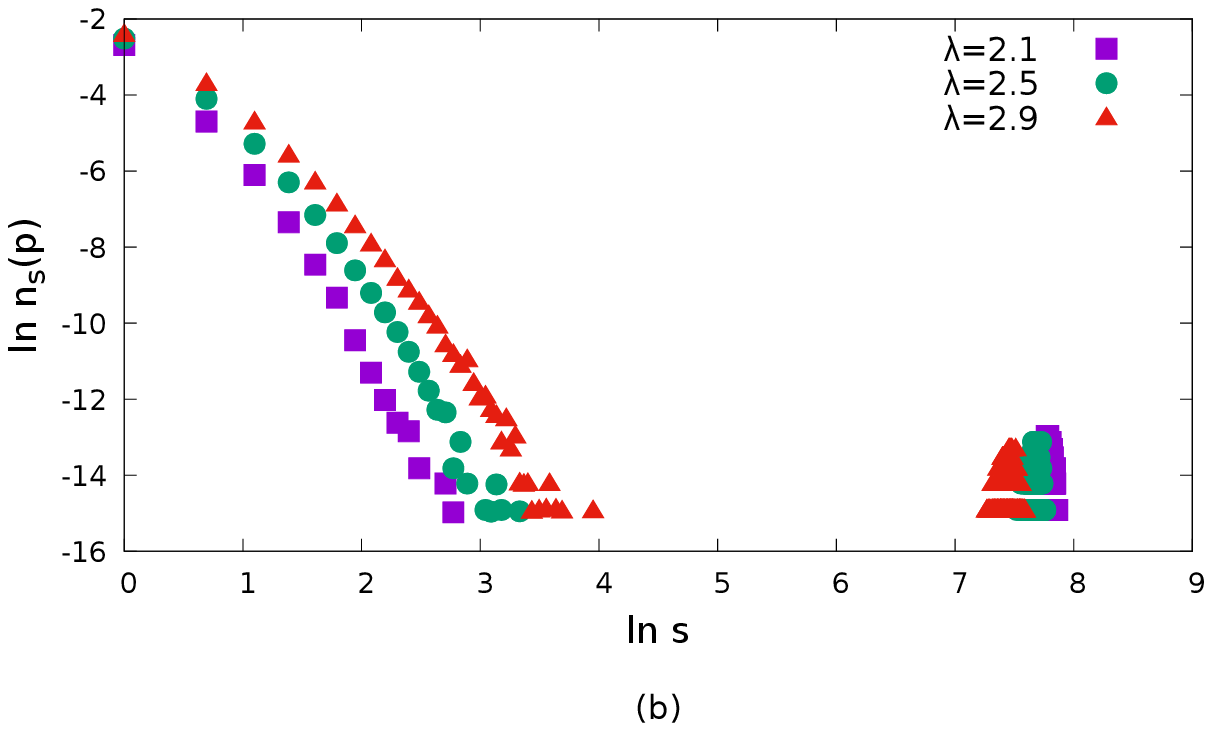}
		\includegraphics[width=70mm]{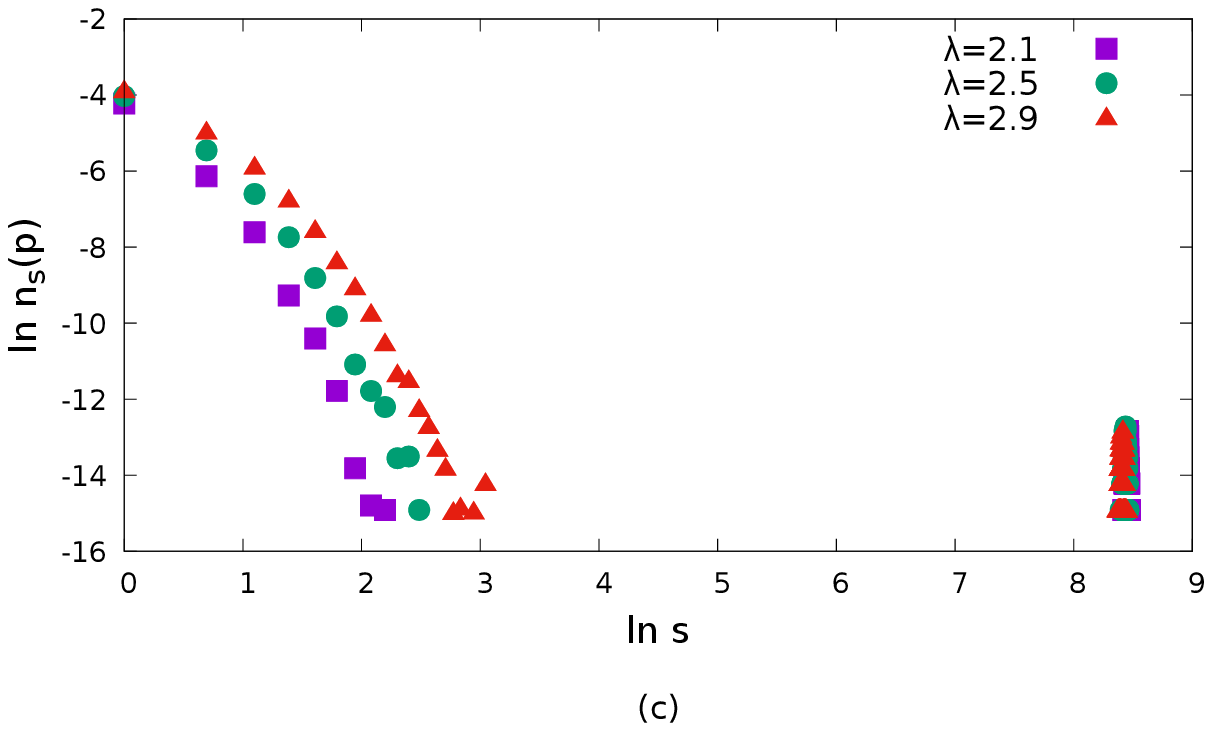}
		\caption{(Color online) Number of clusters per network node  $n_s(p)$, {   averaged over an ensemble of network configurations}, as a function of  cluster size $s$ for a SF network of size 
		$N=6000$ at various $\lambda$ and  concentrations of susceptible
			nodes $p=0.3$ (a), $p=0.5$ (b), $p=0.8$ (c) in a double logarithmic scale. The random scenario  is applied. Probability and a 
			size of the largest component increases both with increasing of $p$
			and decreasing of $\lambda$.}
\label{rozcluster}	
\end{center}
\end{figure}
 
   Our simulation results for the cluster size distribution as averaged over an ensemble of constructed networks are shown in Fig. \ref{rozcluster}.  
   Note that the  percolation thresholds for given values of 
 $\lambda $ are   estimated based on  Eq. (\ref{criterion})  making use of averaged values for $\langle k \rangle(p)$ and 
 $\langle k^2 \rangle(p)$ found in our simulations:
$p_{{\rm perc}}(\lambda=2.1)=0.06$,  $p_{{\rm perc}}(\lambda=2.5)=0.1$, $p_{{\rm perc}}(\lambda=2.9)=0.16$
(since that these are values obtained for a finite system  which are not to be compared with analytical results obtained for infinite networks).
    Both the concentration $p$ of susceptible nodes and 
   the scaling law  parameter $\lambda$ cause essential effects on these distributions. At values of $p$ depicted in Fig.  \ref{rozcluster} the considered 
   networks are above the percolation threshold, and  the giant connected cluster is present in the system.
  At small value of
$p=0.3$, a large amount of 
 clusters of relatively small size can be still found in a system.
 As the value of $p$ increases, the probability to find  small clusters is
decreasing, whereas the fraction of nodes in spanning clusters increases. At $p=0.8$, the probability for a susceptible node to 
 belong to a percolation cluster is very high. 
The fluctuations in sizes of  large clusters emerging in a system in vicinity of percolation threshold
 are caused by finite system size.  On the other hand, the smaller is parameter $\lambda$, the larger is the probability for a node with a very high degree 
 (hub) to be found in the network. The presence of such nodes
 increases the connectivity of  network in the sense, that a larger amount of  nodes belong to the spanning cluster. Indeed, as we can see in  Fig.  \ref{rozcluster},
 at each fixed $p$ the value of
 the probability for a node to belong to largest cluster increases with decreasing $\lambda$ .



\begin{figure}[t!]
	\begin{center}
	        	\includegraphics[width=70mm]{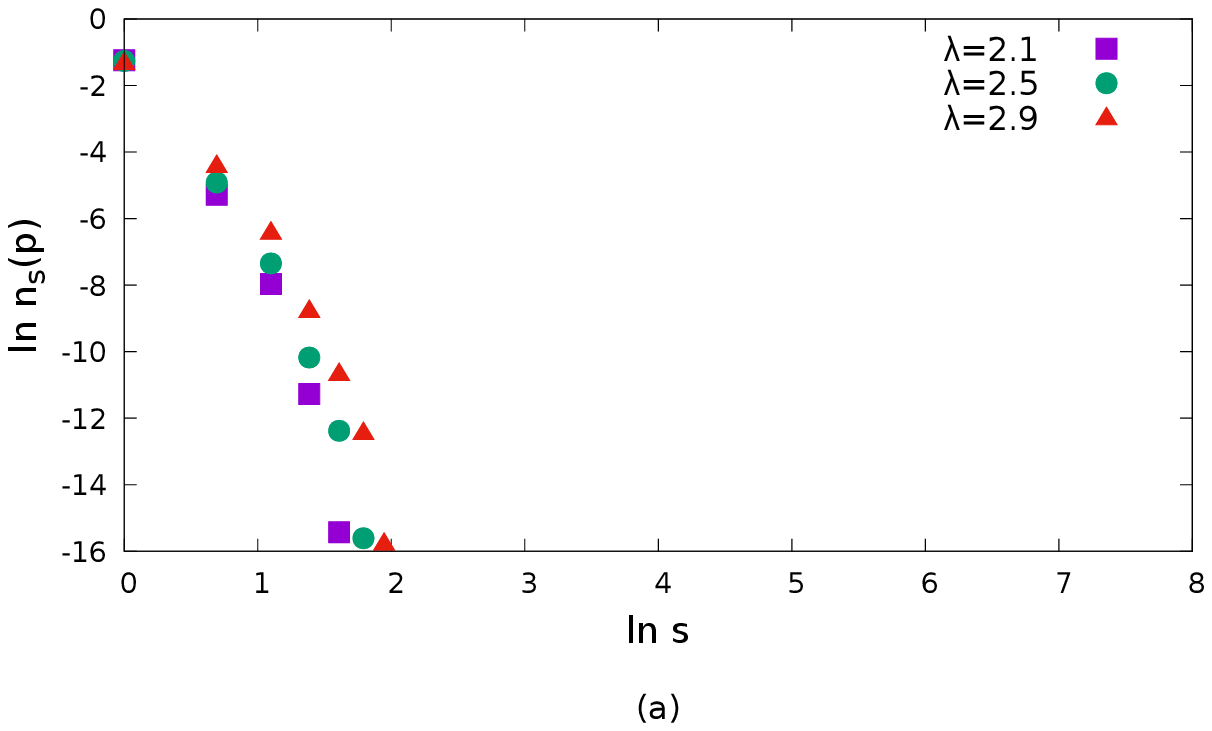}	
	 	\includegraphics[width=70mm]{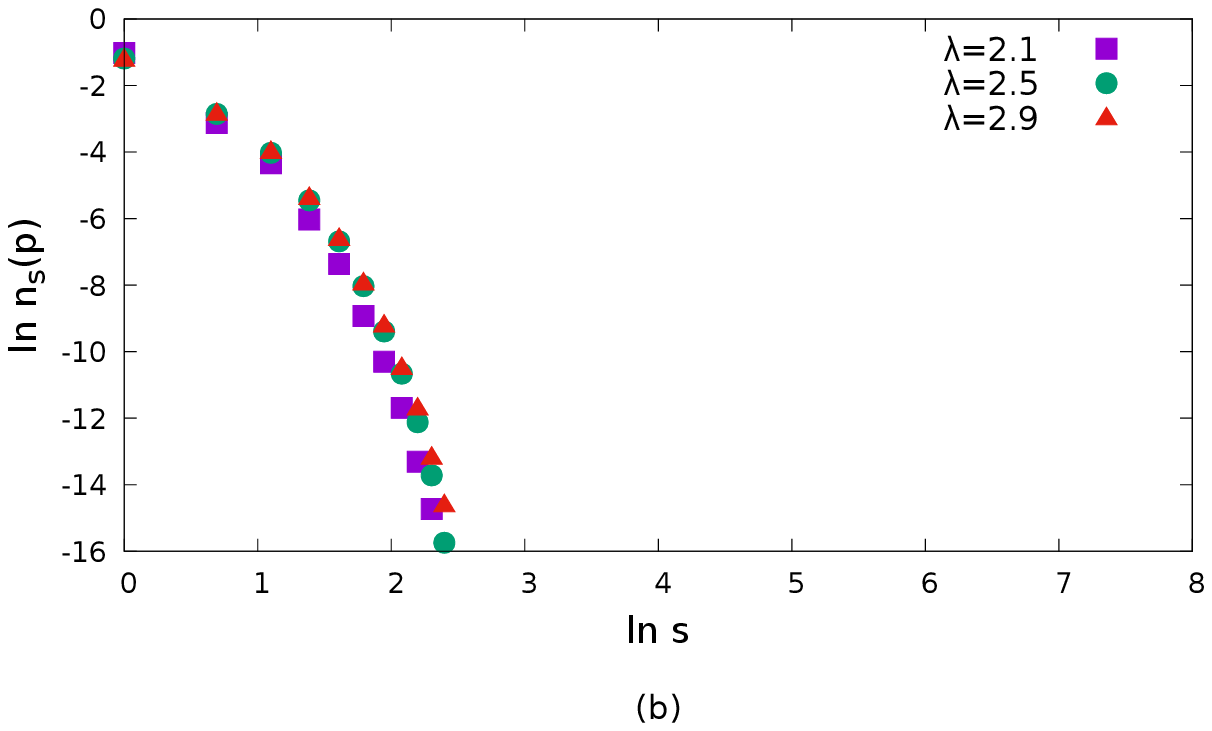}
		\includegraphics[width=70mm]{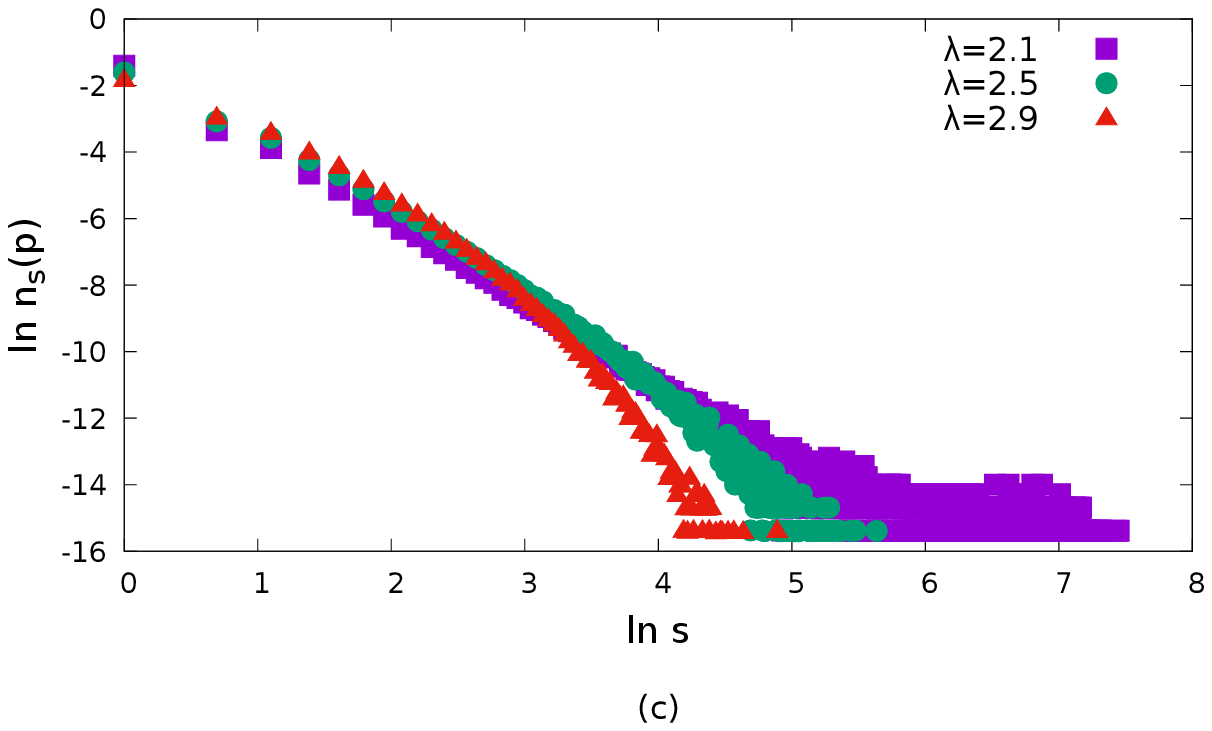}
		\caption{(Color online) Number of clusters per network node  $n_s(p)$, {   averaged over an ensemble of network configurations}, as a function of  cluster size $s$ for a SF network of size 
		$N=6000$ at various $\lambda$ and  concentrations of susceptible
			nodes $p=0.3$ (a), $p=0.5$ (b), $p=0.8$ (c) in a double logarithmic scale. The  {targeted}  scenario  is applied. 
			The large connected component is absent for the low concentrations of susceptible nodes.}
\label{rozclusterhub}	
\end{center}
\end{figure}
 
 \subsubsection{ {Targeted} scenario}

 {In the second, targeted scenario (also called an intentional one, related to the so-called intentional or targeted attacks \cite{Cohen01,Berche09,Berche12})}, 
the fraction $1-p$ 
of  the nodes with the highest connectivity is considered as immune  and removed together with their outgoing links.  The node degree
cutoff  reduces in this case to  the new value $k_{{\rm max}}(p)$, which can be estimated from the relation:
  \begin{equation}
(1-p)=\sum_{k=k_{{\rm max}}(p)}^{k_{{\rm max}}} P(k).\label{kmaxp}
\end{equation}
In turn, the probability $\tilde{p}$ that a link of a given node  is leading to a removed node
is given by \cite{Cohen01}
$
\tilde{p}=\sum_{k=k_{{\rm max}}(p)}^{k_{{\rm max}}} kP(k)/{\langle k \rangle},
$
with $\langle k \rangle$ is calculated on the base of the original distribution before the removal of nodes. Thus, as a  result of an  {targeted} scenario we have
 a SF network with the cutoff $k_{{\rm max}}(p)$, where the fraction $\tilde{p}$ of nodes are randomly removed.
  The threshold value of $p_{{\rm perc}}$ in this situation can be found from the condition (\ref{criterion}),
where the averages  $ \langle k \rangle(p)$,  $\langle k^2 \rangle(p)$  are performed on the base of
  the distribution function:
  \begin{equation}
  \tilde{P}(k)=\sum_{m\geq k}^{k_{{\rm max}}(p)} P(m)\frac{m!}{k!(m-k)!}(1-\tilde{p})^k\tilde{p}^{m-k}
  \end{equation}
   with the upper cutoff 
 $k_{{\rm max}}(p)$ estimated from (\ref{kmaxp}).

{   With the simulated networks at hand, the maximal node degree  $k_{{\rm max}}(p)$ follows automatically.
The estimates for percolation thresholds in this case are evaluated based on Eq. (\ref{criterion}) using the  averaged values for $\langle k \rangle (p)$ and 
 $\langle k^2 \rangle(p)$ found in our simulations:}
$p_{{\rm perc}}(\lambda=2.1)=0.76$,  $p_{{\rm perc}}(\lambda=2.5)=0.85$, $p_{{\rm perc}}(\lambda=2.9)=0.91$ (again, we should mention  that these 
finite-size values are not to be compared 
with the analytic results obtained for infinite networks as given e.g.  in \cite{Cohen01}). Note the pronounced increasing of 
percolation threshold due to removing the nodes with high connectivity, which play the main role in keeping the  network structure connected.


The cluster distributions in the case of    {targeted} scenario are shown  in  Fig. \ref{rozclusterhub}, which should be compared with 
Fig. \ref{rozcluster}. At $p=0.3$ and $p=0.5$, presented networks are below the percolation threshold, 
and one observes a large amount of small clusters of susceptible nodes. Only at $p=0.8$, which is close to the percolation threshold of 
the network with $\lambda=2.1$, we observe an increasing probability for a node to be in a large spanning cluster.

Thus, already at this level of analysis one can immediately conclude about the efficiency of the  {targeted} scenario. Indeed, absence 
of large connected clusters of susceptible nodes up to $p=0.8$ indicates a 
severe difficulty for epidemic  to spread  in such system. 

\section{The method} \label{III}

With the constructed vaccinated networks at hand, we are in position to analyze the spreading phenomena on such structures. 
The various spreading scenarios and cellular automaton algorithm which is used to implement these scenarios in a vaccinated network are described below.

\subsection{Spreading scenarios} \label{Spreadingtypes}

Within compartmental models, that serve a paradigm of epidemic quantitative description
\cite{Kermack27,Anderson92},
the system of agents (population) at any stage of  spreading process  is assumed to contain
the  three main classes  of individuals: susceptible S, infected I, and removed R (either by
recovery and acquiring immunity or by death).  
Each  node of the network thus represents an individual,
which can exist in three possible states, and
each  link  corresponds to the  interaction (contact) between the individuals which allows the disease transmission.
Within the mostly simplified SI scenario, once being infected the susceptible  individuals  stay infected and infectious throughout their life. 
 The SIS
model describes the  disease without  immunity, where  individuals can be infected again and again. In this case, the
so-called endemic state can emerge: the ratio of infected individuals  reaches some equilibrium value, and spreading process continues in time
without termination. The SIR model describes the  disease with acquired immunity: in response of being infected and cured a  person becomes resistant 
to a given infection during a life time. This mechanism
leads to possibility of epidemic outbreak.

In both SIS and SIR cases, the effective spreading rate $\beta$ (also known in epidemiology as a basic reproduction number  \cite{Milligan15})  
plays an essential  role in determining the 
course of process. 
An estimate for the  threshold value of $\beta$ for the SIS model on general complex networks  is given by \cite{Pastor02} :
\begin{eqnarray}
\beta_c^{SIS}=\frac{\langle k \rangle}{\langle k^2 \rangle},\label{RSIS}
\end{eqnarray}
where averaging is performed with the corresponding node degree distribution function.
Only when effective spreading rate on a given network exceeds $\beta_c$, the spreading leads to endemic state, whereas at $\beta<\beta_c$ it stops with time. 
In the asymptotic limit of infinitely large SF  networks,  the second moment of the degree distribution (\ref{power})
diverges at $ 2 < \lambda < 3$  and  one finds $\beta_c \to 0$ , thus confirming the fragility of such 
networks for disease spreading. Note however that for smaller size networks  $\langle k^2 \rangle$  has  a huge but finite value, 
defining an effective nonzero threshold due to finite
size effects, as it is usual in  non-equilibrium phase
transitions \cite{Marro99}.
Correspondingly,  for the SIR model the threshold value $\beta_c$, which separates the non-spreading state
 and regime with epidemic outbreak, is given by \cite{Pastor15}:
\begin{eqnarray}
\beta_c^{SIR}=\frac{\langle k \rangle}{\langle k^2 \rangle-\langle k \rangle}.\label{RSIR}
\end{eqnarray}
 Estimate for $\beta_c^{SIR}$ is directly connected with problem of bond percolation on a corresponding network. 

\subsection{Updating algorithm}  
 
 The classical compartment models \cite{Kermack27,Anderson92}  are formulated within the frames of the   differential equations  and  do 
 not take into account the local structure of the system, serving in some sense like a mean-field approximation (each individual is assumed 
 to interact simultaneously with all the other individuals).  
A more advanced description is provided e.g. by  the cellular automaton (CA) means, which is  based on the local properties of the system 
and allows to evaluate the
global properties from the local ones.  Here, one considers
a set of individuals,  each being  attached to a node of an underlying graph. 
 The evolution
of the system is  governed by a specific update algorithm,  which at each discrete moment of time 
changes the state of  individuals according
to the states of their  neighbours in the graph.
 
  We  consider the  synchronous version of CA updating algorithm,  {where one time step
implies a sweep throughout the whole system  \cite{Grassberger83}.
 At time $t$, the $k$th node is considered to be in a state $\sigma_k(t)$, with $\sigma_k(t)$ taking values from a set  $\{0,1,2\}$,
corresponding to {\em S}, {\em I}, or {\em R}  {respectively}.  Let $S(t)$, $I(t)$ and $R(t)$ be fractions   of healthy (susceptible), 
infected and recovered individuals, located on the nodes of the network, such that
$S(t)+I(t)+R(t)=1$. 
Thus, the global fractions  {\em S}, {\em I}, and {\em R} are given by:
\begin{eqnarray}
&&S(t)=\frac{1}{N}\sum_{k=1}^N\delta_{0\sigma_k(t)},\nonumber\\
 &&I(t)=\frac{1}{N}\sum_{k=1}^N\delta_{1\sigma_k(t)},\nonumber\\
 && R(t)=\frac{1}{N}\sum_{k=1}^N\delta_{2\sigma_k(t)},\nonumber
\end{eqnarray}
where $\delta$ is the Kronecker delta, $N$ is the total number of agents 
 {and summation is performed over all network nodes.
To further account for the heterogeneity of a SF network, it has been suggested
to consider the probability of a local short-distance disease
transmission differing from the spreading rate $\beta$ \cite{Gagliardi10}:}
\begin{equation}
p_k(t)=1-(1-\beta)^{n_k(t)}, \label{nn}
\end{equation}
where  $n_k(t)$ is a number of  infected nearest neigbours of a node $k$ at time $t$.
   The time update $t-1 \to t$ consecutively considers all $k=1, \dots, N$
nodes  and changes each state $\sigma_k(t-1) \to \sigma_k(t)$ according to the rules described below.


{\em (i) SI model}
\begin{itemize}
	\item  {Choose node $k$.}
	\item If $\sigma_k(t-1)=1$,  {then do nothing} so that  $\sigma_k(t)=1$.
\item  If $\sigma_k(t-1)=0$, then the state is changed to 1 with probability $p_k(t-1)$ (Eq. (\ref{nn})), so that $\sigma_k(t)=1$.
\end{itemize}

{\em (ii) SIS model}

\begin{itemize}
	\item  {Choose node $k$.}
	\item  If $\sigma_k(t-1)=1$, then it is cured with probability $1$ so that $ \sigma_k(t)=0$.
  \item  If $\sigma_k(t-1)=0$, then the state is changed to 1 with probability $p_k(t-1)$ (Eq. (\ref{nn})), so that $\sigma_k(t)=1$.
\end{itemize}

{\em (iii) SIR model}
\begin{itemize}
	\item  {Choose node $k$.}
	\item  If $\sigma_k(t-1)=2$,  then do nothing.
  \item If $\sigma_k(t-1)=1$, then with probability $1$ it is recovered, so that $\sigma_k(t)=2$.
  \item If $\sigma_k(t-1)=0$, then the state is changed to 1 with probability $p_k(t-1)$ (Eq. (\ref{nn})), so that $\sigma_k(t)=1$.

\end{itemize}
 
As already mentioned above, we will analyze spreading processes in the case, when only a  selected fraction $p$ of
 nodes (agents) is susceptible to disease. To this end, applying the rules of random or  {targeted} scenario,  at initial moment of time $t=0$ we 
 consider each node $k$ of the network to be either susceptible to disease with
 probability $p$
or immune with  probability $1-p$ (the state of corresponding nodes  $\sigma_k=0$ always, and these nodes are considered as
 non-active in an updating algorithm).

The maximum number of time steps is taken $t=400$ (typical times for  reaching the
stationary state for the cases studied below are found  in general at $t<100$).


\section{Results} \label{IV}

We apply the cellular automaton mechanism to study the epidemic process on networks with fraction $p$ of active nodes, obtained as a result of both 
random and  {targeted} scenarios, as described in the previous Section. 
At time $t=0$,  we chose randomly a small fraction $i_0=0.01$ of susceptible nodes, which are supposed to be  infected.  The 
 disease  spreading process from infected to susceptible agents is started then.

\subsection{SI model}

Since the equilibrium values of 
$I^*(p)$ and $S^*(p)$ do not depend on $\beta$ in the case of this scenario,
 we fix the value for the infection rate $\beta=0.5$   
to make the evolution process  fast.
One can easily estimate the fraction of nodes  $S^*(p)$
 which remain susceptible and not affected by the spreading process, using the arguments from our previous work on the spreading on regular lattice 
 \cite{Blavatska21}. Indeed, 
 if at $t = 0$ in some cluster of size $s$ of susceptible agents no
one node gets infected, this cluster will remain safe till the epidemic terminates (this cluster will not be touched by the
epidemic process in any way).
   $S^*(p)$ thus equals to the fraction of nodes $ P_{{\rm safe}}(p)$ in these ``safe clusters'', which can be easily estimated as:
 \begin{equation}
 S^*(p)=P_{{\rm safe}}(p)=\sum_{s}(1-i_0)^sP_s(p). \label{psafe}
\end{equation}
 {Here, $P_s(p)$ is the probability to find a cluster containing $s$ susceptible nodes in a network vaccinated according to given scenario (as 
 discussed in Section \ref{rozpodilSec}),
  and the sum  spans over all available clusters of susceptible nodes.} The equilibrium value for the fraction of infected agents is thus 
  straightforwardly given by $I^*(p)=p- S^*(p)$.

 \begin{figure}[t!]
	\begin{center}
		\includegraphics[width=80mm]{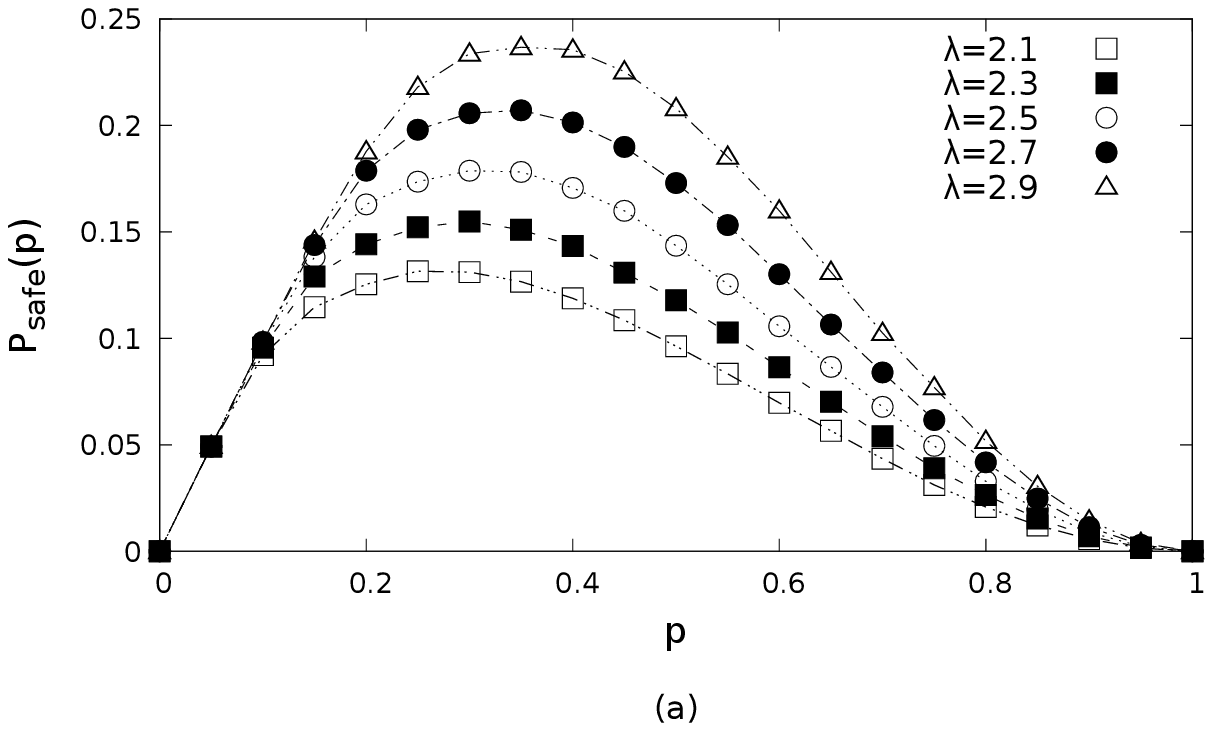}
		\includegraphics[width=80mm]{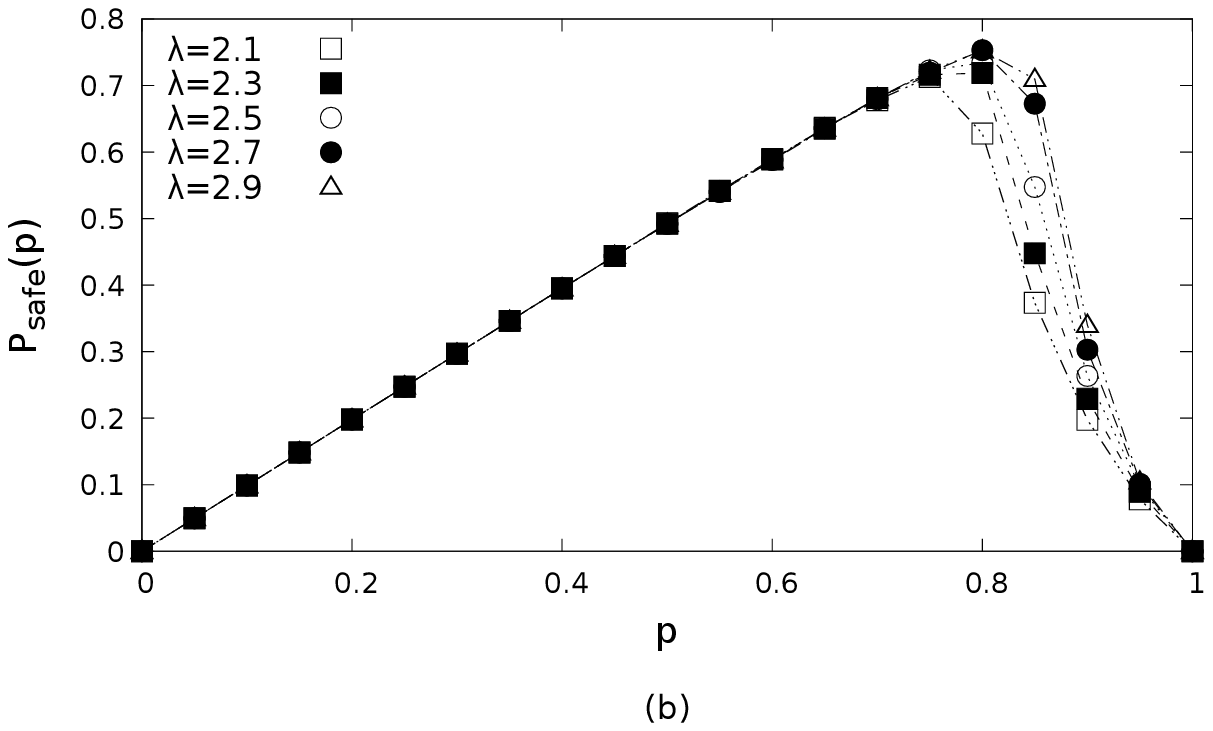}
		\caption{ \label{psi}
		Numerical values for  $P_{\rm safe}(p)$ (equal to equilibrium values of fraction of susceptible nodes $S^*(p)$ in SI model)  as a
		function of $p$ at different $\lambda$ in random scenario (a) and  {targeted} scenario (b). The maximum number of individuals 
		not affected by spreading process increases with decreasing the parameter $\lambda$.}
	\end{center}
\end{figure}
  
 The data for  $S^*(p)$ obtained 
 in our simulations as results of random and  {targeted} vaccination  are presented in Fig. \ref{psi}. These results can be treated as a direct 
 consequence of performed vaccination, which  is not only 
protecting  immunized individuals, but also increases the fraction of susceptible 
individuals not affected by disease. Note, that for each $\lambda$, in both scenarios we obtain a maximum
 of unaffected individuals at some value of $p_{{\rm max}}$, which can be explained based on the same considerations as 
 presented in our previous analysis of regular lattices \cite{Blavatska21}.  {To this end, let} us recall the cluster size distributions as 
 presented in Figs. \ref{rozcluster} and \ref{rozclusterhub}. 
  At small values of $p$ (up to some critical one $p_{{\rm max}}$), there is a large amount of clusters of small size, and “diversity” of clusters 
  (number of different clusters)  increases with $p$.
  Thus,  the probability for one of them to get infected gets lower and that is why $P_{{\rm safe}}(p)$  increases in this region. On the other hand, above the 
  value of  $p_{{\rm max}}$  the number of various clusters decreases (small
clusters start to segregate into the larger ones), which in turn
makes it easier for infection to  {propagate} through the system (thus also  $P_{{\rm safe}}(p)$ starts to decrease). For the networks under consideration, 
in the case of random scenario $p_{{\rm max}}$  is found to be around $0.3$,
whereas for the case of  {targeted} scenario it is observed to be around $0.8$.

 \begin{figure}[b!]
 	\begin{center}        	
	 	\includegraphics[width=80mm]{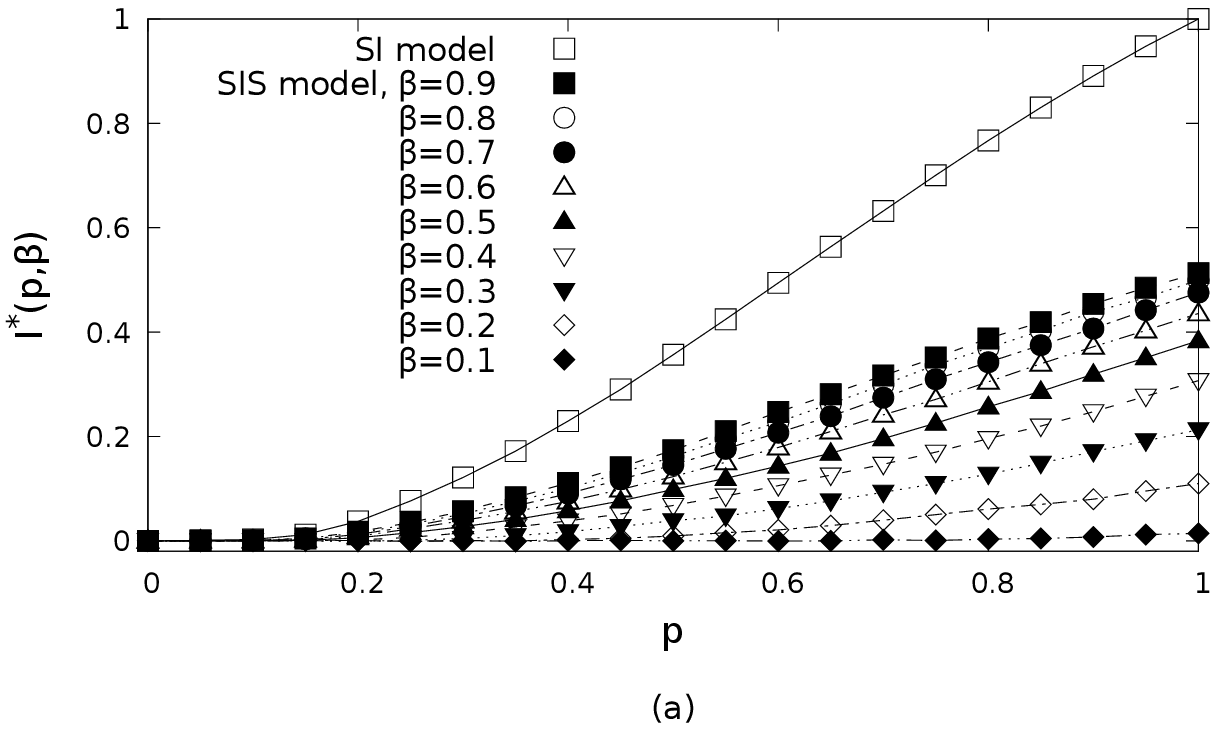}
	 	\includegraphics[width=80mm]{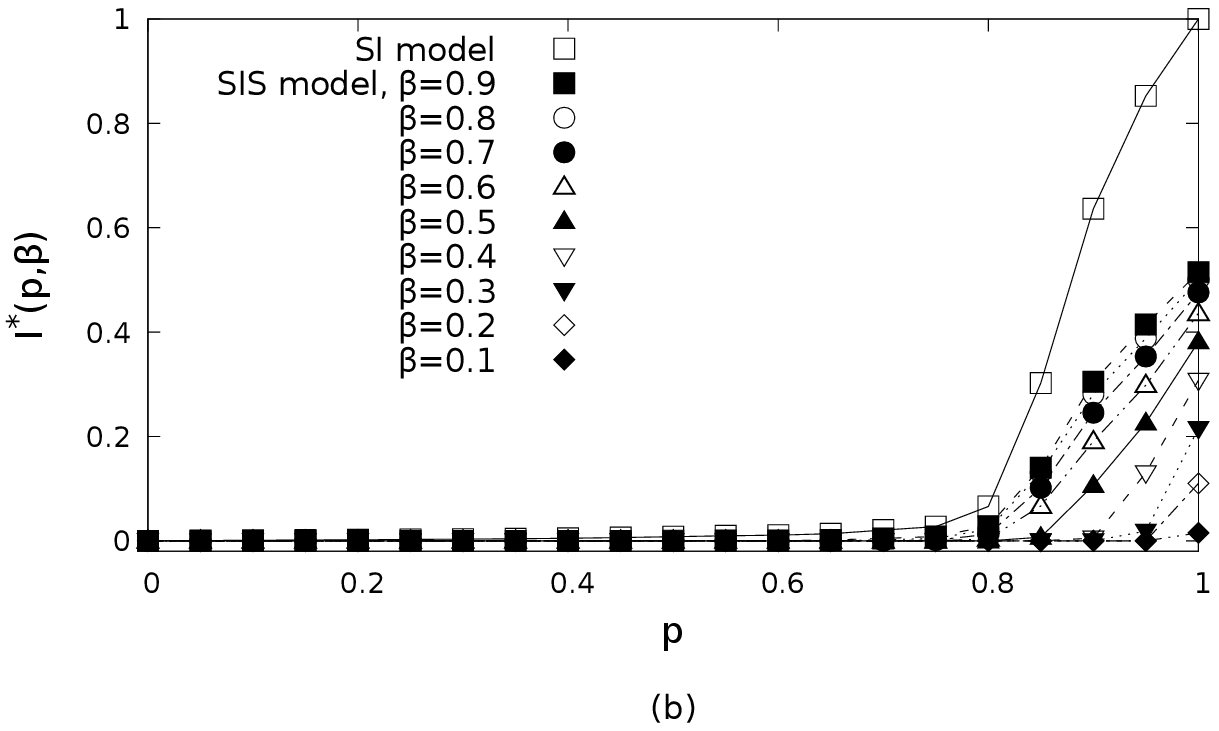}	
		\caption{Total fraction of infected nodes in equilibrium as function of $p$ for a SF network
		with $\lambda=2.5$ for the SI (open squares) and SIS (all other symbols at different $\beta$) models.  Lines are guides to the eye.  
		a):  random scenario, b):  {targeted} scenario.}
		\label{infSIS}	
\end{center}
\end{figure}

Note, that in the case of random scenario,  the number of  unaffected nodes  strongly depends also on the 
parameter $\lambda$. Again, recalling the cluster distribution  as shown on Fig. \ref{rozcluster}, the smaller the parameter $\lambda$,  the larger the 
probability for a big cluster of susceptible nodes to occur, which makes the system more
fragile for infection spreading. Indeed,  $S^*(p)$ decreases with decreasing $\lambda$. The situation is very different in the case of the
 {targeted} scenario, where there is practically no difference 
 in $P_{{\rm safe}}(p)$ in the case of strong vaccination (for $p$ below 80\%). It can be explained by the fact, that the main qualitative 
 difference between networks with different $\lambda$ is caused by presence of hubs, which are stronger and 
 make the system more connected, the smaller is parameter $\lambda$.   Removing the hubs from spreading process by vaccinating them leads to 
 breaking the connectivity of network , which now consists of a large number of disconnected clusters of small sizes, as 
 shown in Fig. \ref{rozcluster}. In this respect, there is practically no difference in cluster distribution at various $\lambda$.
Comparing the data  of Fig. \ref{psi}, one immediately notices the effectiveness of  {targeted} vaccination. 
Indeed, in this case the maximum number of individuals, not affected by disease, reaches almost 80\% of population at  
corresponding $p_{{\rm max}}$ value and is almost independent on  $\lambda$,
whereas in the case of random vaccination it does not exceed 25\%.

\subsection{SIS model}

 \begin{figure}[t!]
 	\begin{center}        	
	 	\includegraphics[width=80mm]{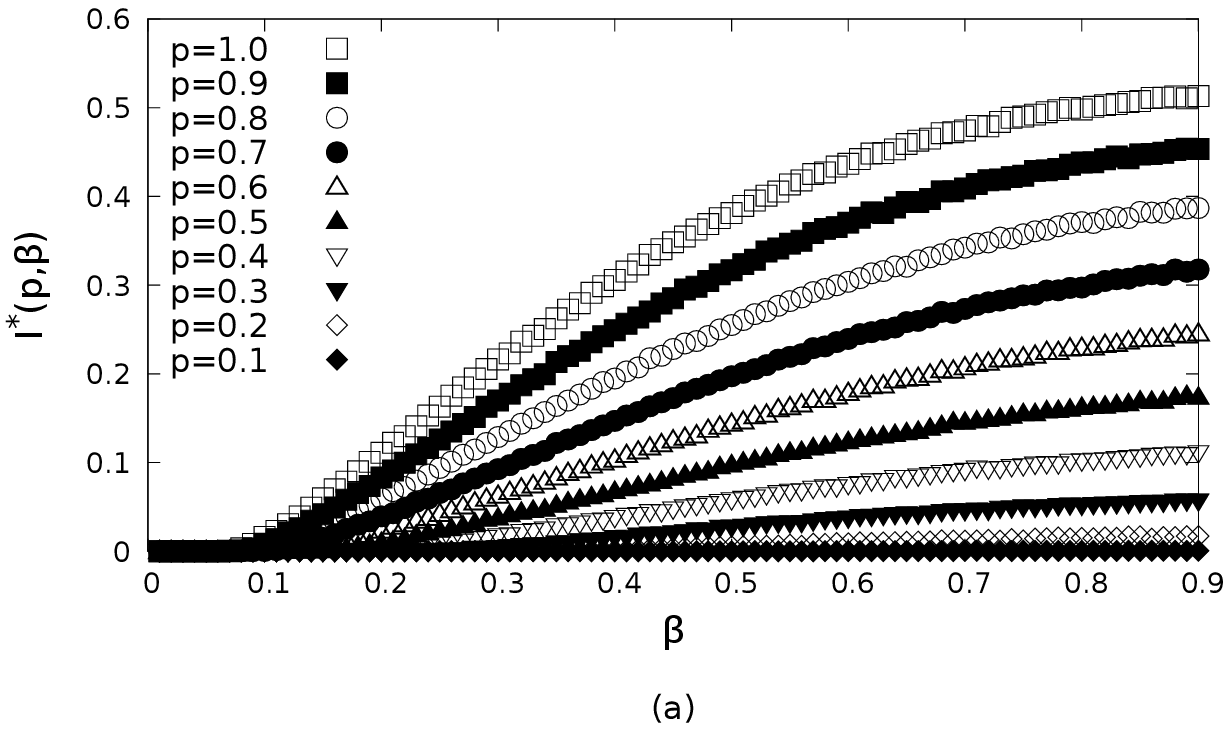}
	 	\includegraphics[width=80mm]{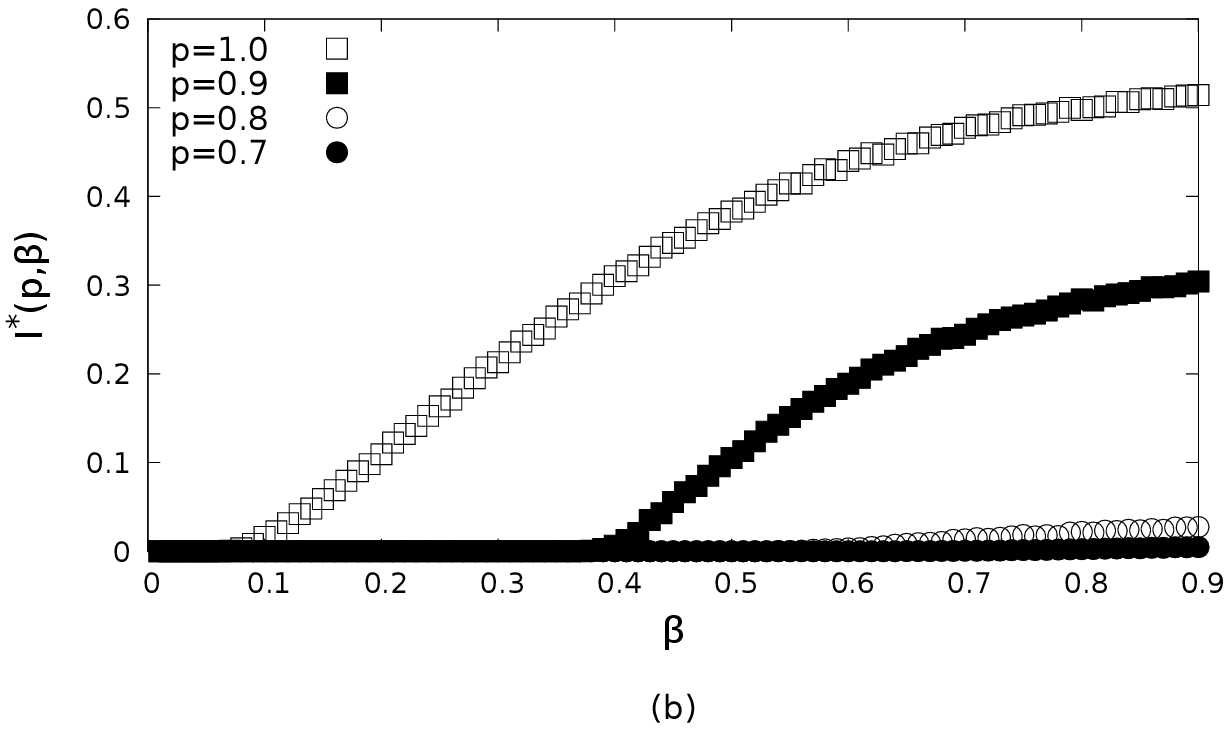}	
		\caption{ Total fraction of infected nodes in equilibrium for a SF network with $\lambda=2.5$ for the SIS model as function of $\beta$.   a):  
		random scenario, b):  {targeted} scenario.}
		\label{infSIScrit}	
\end{center}
\end{figure}

Within the frames of SIS scenario, the infected individuals get cured without obtaining  immunity, so that they can be infected over and
over again, undergoing a cycle $S \to I \to S$. After reaching the equilibrium state, one obtains the values $S^*(p,\beta)$ and  $I^*(p,\beta)$ for 
the fractions of susceptible and infected individuals, which now are defined not 
only by the initial amount of susceptible individuals $p$, but also by the spreading rate $\beta$.

 \begin{figure}[b!]
 	\begin{center}        	
	 	\includegraphics[width=80mm]{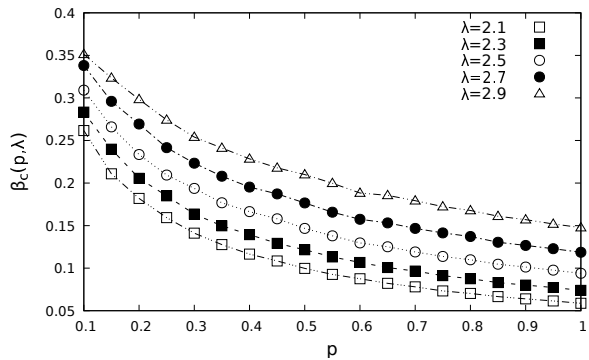}
		\caption{Critical value of spreading rate $\beta_c(p,\lambda)$  for a SF network in SIS model in a random scenario.}
		\label{betaSIS}	
\end{center}
\end{figure}

We start  with analyzing the averaged amount of infected nodes,  which can be observed in a network in equilibrium state. 
 In Fig. \ref{infSIS} we give the example of our numerical estimates for the case $\lambda=2.5$,  applying both the random and  {targeted} scenarios of vaccination.
   As expected, $I^*(p,\beta)$ in both cases is considerably smaller compared to the simplified
 SI model  and decreases with lowering the spreading rate $\beta$. These data allow us to directly compare the effectiveness of two types of vaccinations. 
 Let us recall that 
 the minimum fraction  $p_{c}(\beta)$ of susceptible individuals, at which the infection can ``survive' in a system (the value of $I^*(p,\beta)$ is above zero),
 can be estimated as  \cite{Pastor02b}:
 \begin{equation}
 \frac{\langle k \rangle(p=p_c)}{ \langle k^2 \rangle(p=p_c)}={\beta}, \label{minvacint}
\end{equation}
 where the averages $\langle k\rangle (p)$, $\langle k^2 \rangle(p)$ are performed with the degree distribution  of the network resulting after the removal of 
 the $(1-p)N$ nodes of highest degree.
 
 Making use of the data for $\langle k \rangle$, $\langle k^2 \rangle$ obtained in our
 simulations, we have for example at $\lambda=2.5$:    $p_c(\beta=0.6)=0.15$, $p_c(\beta=0.3)=0.3$, $p_c(\beta=0.1)=0.91$. For the case of  {targeted} 
 vaccination, making use of Eq. (\ref{minvacint})
 gives us corresponding  estimates:  $p_c(\beta=0.6)=0.75$, $p_c(\beta=0.3)=0.85$, $p_c(\beta=0.1)=1.0$, so that 
 one observes a non-zero value of $I^*(p,\beta)$  at considerably larger values of $p$  comparing to the random scenario.
 One observes  no spreading activity ($I^*(\beta,p)$ is practically zero) in this case at $p<0.8$ even at very strong disease spreading  {rate} (large $\beta$).     
 Such a quantitative difference between two scenarios can be intuitively understood, recalling the  peculiarities of distribution of clusters of susceptible nodes  
 (as described in Subsection \ref{rozpodilSec}). Indeed, at the same value of  $p$,
  {targeted} scenario leads to formation of a large amount of separated clusters of very small size, as compared with the random scenario. This leads to 
 restriction of infection in very small separated regions
  and preventing its spreading around in spanning clusters, as it takes  place during the random vaccination scenario.


Analysis of system behaviour at various values of parameter $\beta$ reveals 
a kind of phase transition  between the so-called endemic state (with non-zero equilibrium value of $I^*(p, \beta)$ ) and disease-free states, where the role of
the order parameter is played by the stationary value of  $I^*(p, \beta)$ .
The phase diagrams of corresponding  phase transitions 
are given in Fig. \ref{infSIScrit}.
  The critical value of the 
effective spreading rate $\beta_c(p,\lambda)$ is defined as those separating
 infected and healthy phases.
  We obtained the numerical estimates for $\beta_c(p,\lambda)$ by making use of 
  fitting of data for $I^*(p, \beta)$ to the form $I^*(p, \beta)=A(\beta_c(p,\beta)-\beta)^{\gamma}$, where $A$ is  constant.
  The
  results are presented in Fig. \ref{betaSIS}.  Note that at $p=1$ the results for $\beta_c$ are in a good correspondence with estimates 
  based on Eq. (\ref{RSIS}) and with the numerical simulations performed in  \cite{Ferreira12}.  
  Note that we present the results for  $\beta_c$ only for the case of random vaccination scenario, since for the  {targeted} scenario for $p$ above $0.8$  
  practically  no infection activity  is observed in system even at very high values of 
spreading  {rate}  $\beta$.

 \begin{figure}[b!]
 	\begin{center}        	
	 	\includegraphics[width=80mm]{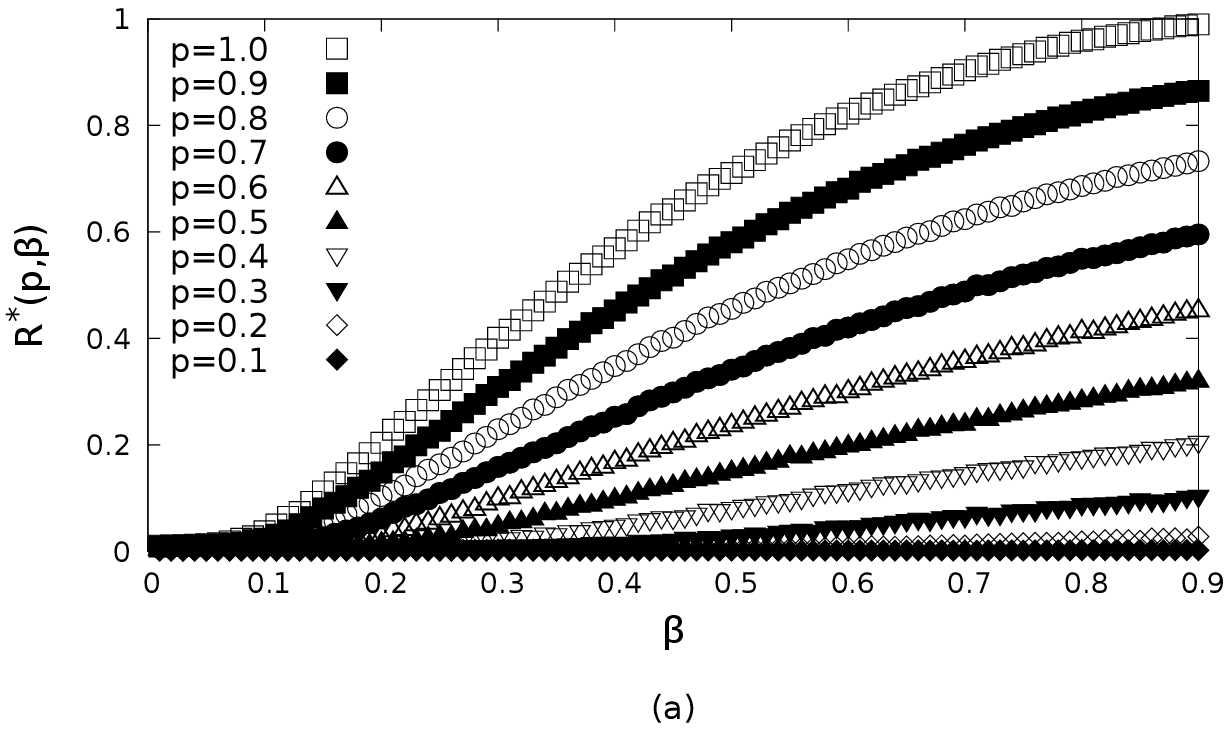}
	 	\includegraphics[width=80mm]{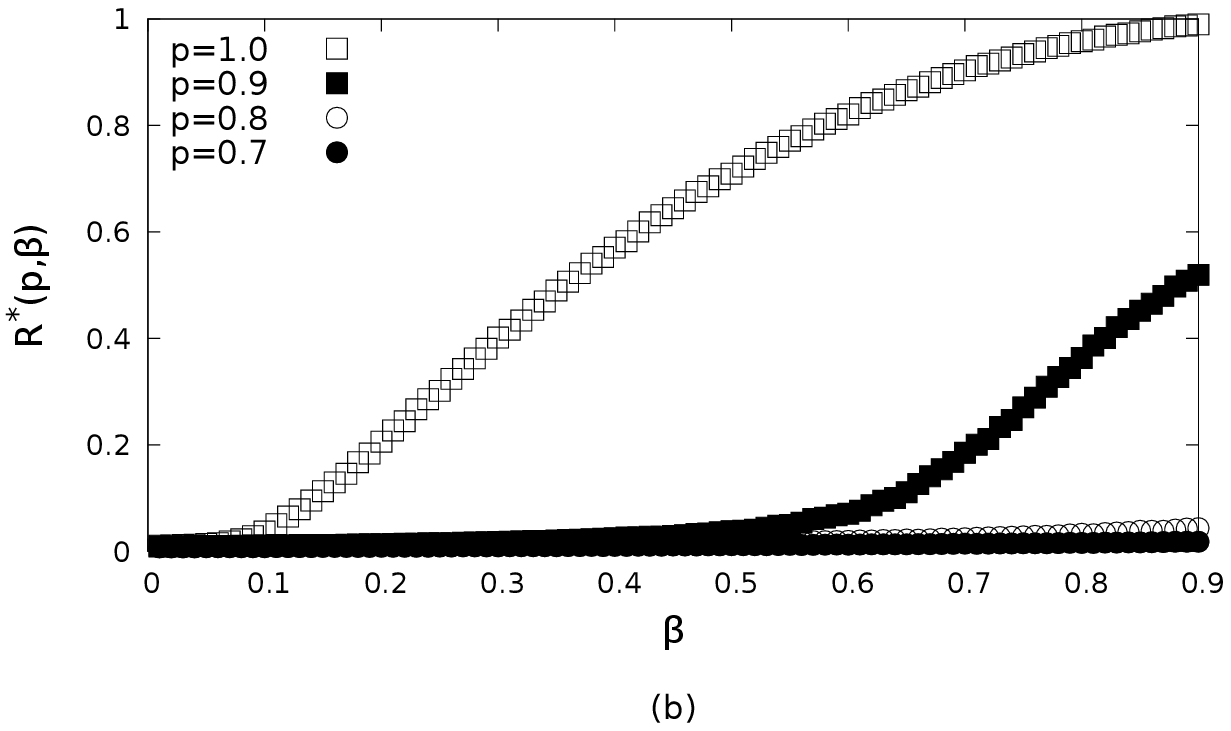}	
		\caption{Total fraction of removed nodes in equilibrium for a SF network with $\lambda=2.5$ for the SIR model as a function of $\beta$.  a):  
		random scenario, b):  {targeted} scenario.}
		\label{infSIRcrit}	
\end{center}
\end{figure}

As expected,  the threshold value of $\beta_c(p,\lambda)$ increases with decreasing parameter $p$: infection should be ``stronger'' 
to occupy the system, if fraction of vaccinated individuals increases. On the other hand, 
$\beta_c(p,\lambda)$ decreases with decreasing the parameter $\lambda$. It is connected with the fact, that the smaller the $\lambda$, the 
larger is the probability to find high-degree nodes in the network. This makes  the spreading of infection on such networks much more easier.

\subsection{SIR model}


Within the frames of SIR scenario, which can be described by a simple scheme $S \to I \to R$,
the infected individuals get immune, 
so that the repeated infection  {of the same node} is not allowed in the course of  such process. After reaching the equilibrium state, 
the fraction $R^*(p,  \beta)$ of individuals
 becoming removed (or with the life-long immunity) is observed (obviously $I^*(p,  \beta)$ is always zero in this scenario, independently on details of the 
 epidemic spreading).
The rest of individuals ($S^*(p,  \beta)=p-R^*(p,  \beta)$) remain susceptible. 


 \begin{figure}[t!]
 	\begin{center}        	
	 	\includegraphics[width=80mm]{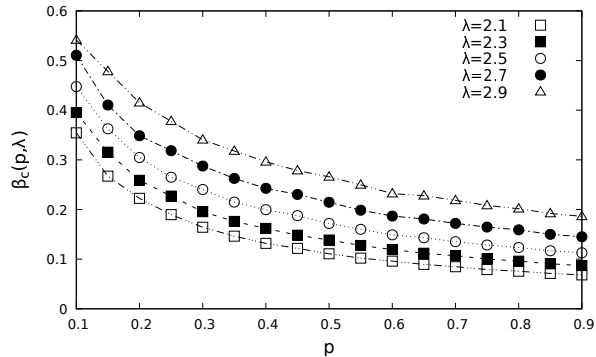}
		\caption{Critical value of spreading rate $\beta_c(p,\lambda)$  for a SF network in SIR model in a random scenario.}
		\label{betaSIR}	
\end{center}
\end{figure}

 The SIR model also exhibits a transition between a phase
where the disease outbreak reaches a finite fraction of the
population and a phase where only a limited number of
individuals is affected. The
order parameter is therefore the density of removed 
 individuals in an equilibrium state, which is defined by the
effective spreading rate $\beta_c$, separating
 infected and healthy phase.
The phase diagrams of corresponding  phase transitions 
are given in Fig. \ref{infSIRcrit}.
Qualitatively, the picture resembles the transition between endemic and healthy states within the frames of SIS model (Fig. \ref{infSIScrit}). 
Again,  for the case of  {targeted} vaccination
 one observes a non-zero value of $R^*(p,\beta)$  at considerably larger values of $p$  comparing with the random scenario, and there is
  no spreading activity ($R^*(\beta,p)$ is practically zero) in this case at $p<0.8$ even at very strong disease spreading  {rate} (large $\beta$).     

  Our numerical estimates for $\beta_c(p,\lambda)$ for the case of random scenario are presented  in Fig. \ref{betaSIR}.  
 At $p=1$ the results for $\beta_c$ are in a good correspondence with the estimates based on Eq. (\ref{RSIR}). Similarly to the case of SIS scenario, 
 the threshold values of $\beta_c(p,\lambda)$ increase with decreasing parameter $p$ and decrease with decreasing  $\lambda$. 
Note also, that at each $p$ and $\lambda$ values  the epidemic spreading for the SIS model occurs
at a smaller value of  $\beta_c$ comparing with the SIR model. In the latter case,
the susceptible individuals  cannot be infected
multiple times, which effectively makes the spreading process more weak as comparing with the SIS scenario.

\section{Conclusions} \label{V}

The goal of our study was to reveal peculiarities of infection  spreading in the
environment comprising agents of two types: those that can be a subject of infection
and those that do not participate in the spreading, being ``immune''.
To this end, we considered a  model
in which only some fraction $p$ of agents-nodes
are  susceptible to the disease transmission, whereas the other $1-p$ are treated as immune (vaccinated).
In this respect, the present work continues our previous study \cite{Blavatska21}, where the similar situation was 
analyzed on the background of a regular square lattice.  One of the crucial points there was the emergence of the 
so-called 
safety patterns of susceptible sites, surrounded by immune ones, which remain unaffected by disease
until the stationary state is reached.  Detailed analysis of safety pattern distribution   thus enables one to determine 
the fraction of infected agents in a stationary state. In the present study we aim to extend this analysis to the case of 
a more complex underlying structures. To this end, it is natural to choose the SF networks, which are more closely related 
to real world situations. The question of interest is the  impact of local heterogeneity of such structures on the 
safety pattern distribution and thus on the peculiarities of spreading processes in such systems.

With the said above in mind, we have modeled the spreading process on SF networks with node degree distribution 
given by Eq. (\ref{power})   with the decay exponent $\lambda$ ranging within $2 < \lambda <3$. The dynamics of  
spreading phenomena on such structures is of interest in many aspects,  examples ranging from computer
virus infections  to epidemic diseases. Due to the heterogeneity of SF networks, two  immunization scenarios have been applied.  
In the simplest  random immunization (as exploited in our previous study of a regular lattice),
a share of nodes is randomly 
chosen and made immune.  More effective  targeted scenario implies the targeting the highest degree nodes. To generate 
vaccinated networks,  we made use of the so-called configuration
model.  

It is worth noting that  the effects under consideration cannot be studied within traditional 
analysis of compartmental models based on {the homogeneous mixing hypothesis and the differential equation 
approach \cite{Kermack27}}. There,   the local structure is
 excluded from consideration and only the global macroscopic features can be catched. On the other hand,
 the microscopic analytic treatment is highly complicated due to the strong heterogeneity of the agent-based model under 
 consideration. In such situation, it is optimal to use the cellular automaton approach, based on thorough taking 
into account details of local interactions between agents
on the microscopic scale. Within this approach, we have implemented  three basic spreading scenarios (SI,
SIS and SIR)  for variable ratio $ (0\leq p\leq1)$ of susceptible nodes. 
We studied the susceptible clusters (the safety patterns) distributions in the case of both random and   targeted scenarios.
It has been analyzed, how the  removing of even the small fraction of the highest degree nodes  within the latter scheme 
results in corrupting the spanning clusters and thus preventing the infection spreading.  The main impact of the heterogeneous 
structure of the underlying network manifests itself already on this level of analysis. 
Whereas for the random scenario the fraction of nodes in safety patterns in a SF network is qualitatively comparable 
with those of the regular network,
the targeted scenario leads to the pronounced increase of the number of such nodes (cf. Fig. \ref{psi} of the present study 
and Fig. 3 of Ref. \cite{Blavatska21}).  Already within the  frames of simplified SI model, one notices the effectiveness of  
a targeted vaccination: e.g. at $\lambda=2.5$ the maximal amount of  individuals, not affected by disease, reaches almost 
80\% of population. Note that in the case of random vaccination it does not exceed 25\%. 
Similar picture was observed while analyzing SIS and SIR models: within the  targeted scenario, the  spreading process was 
shown to be prevented  at $p<0.8$ even at very strong disease spreading rate (large $\beta$).    

  The numerical estimates for the  threshold values of spreading parameters $\beta_c(p,\lambda)$  have been obtained 
  within SIS and SIR models within the frames of random 
  scenario. In both cases,  
  $\beta_c(p,\lambda)$ increase with decreasing parameter $p$. This is intuitively understood since 
   when the fraction of vaccinated individuals increases, the infection has to be ``stronger''  to occupy the whole system.  
   On the other hand, the heterogeneity of the SF network, which gest more pronounced with an increase of $\lambda$, leads 
   to decrease of  the threshold value of $\beta_c(p,\lambda)$ (as shown in 
   Fig. \ref{betaSIS}). Again, one can compare  these results with the corresponding values obtained in our previous study  
   (Fig. 11 of Ref. \cite{Blavatska21}). This leads to the conclusion that within a random scenario, a SF network is more 
   compliant for disease spreading as comparing with the regular lattice structure, and 
   this effect is more pronounced with increasing the parameter $\lambda$. On the other hand, vaccination within the 
   targeted scenario makes SF networks incomparably more resistant to epidemic spreading than the regular lattice structures.

\section*{Acknowledgments}

This work was supported in part by National Research
Foundation of Ukraine, project “Science for Safety of Human and Society” No. 2020.01/0338 (VB) and 
 by the National Academy of Sciences of Ukraine, project KPKBK 6541030 (YuH).


\begin{thebibliography}{50}


\bibitem{Blavatska21}
V. Blavatska and Yu. Holovatch, Physica A {\bf 573}, 125980 (2021)


\bibitem{Moore01}
C. Moore and M. E. J. Newman, Phys. Rev. E  {\bf  61}, 5678
(2000)

\bibitem{Murray88}
W.H. Murray, Computers and Security {\bf 7}, 139  (1988)

 \bibitem{Newmann02}
 M.E.J. Newman, S. Forrest, and J. Balthrop, Phys. Rev. E {\bf  66},
035101(R) (2002)

\bibitem{Daley65}
 D.J. Daley and D.G. Kendal,  J. Inst. Maths Applics {\bf 1}, 42 (1965) 


 \bibitem{Murray93}
 J.D. Murray,  {\it Mathematical Biology} (Springer Verlag,
Berlin, 1993)

\bibitem{Barabasi99}
A.-L. Barab\'asi and  R. Albert, Science {\bf 286}, 509 (1999)

\bibitem{Faloutsos99}
M. Faloutsos, P. Faloutsos,  and C. Faloutsos,  Comput. Commun. Rev. {\bf 29}, 251 (1999)

\bibitem{Caldarelli00}
G. Caldarelli, R. Marchetti, and L. Pietronero, Euro-phys. Lett. {\bf 52}, 386 (2000)




\bibitem{Goh02}
K.-I. Goh, B. Kahng, and D. Kim. Phys. Rev. Lett {\bf 88}, 108701 (2002)

\bibitem{Newman01}
M.E. Newman,  Phys. Review E {\bf 64},   016131 (2001)

\bibitem{Navia16}
A. F. Naviaa,  V. H. Cruz-Escalonac, A. Giraldob, and A. Barausse, Ecol. Modell. {\bf 328}, 23  (2016) 

\bibitem{Kermack27}
W.O. Kermack and A.G. McKendrick, Proc. R. Soc. Lond.
Ser. A Math. Phys. Eng. Sci. {\bf 115},  700 (1927)

\bibitem{Anderson92}
R.M. Anderson and R. M. May, {\it Infectious Diseases in
Humans} (Oxford University Press, Oxford, 1992)

 \bibitem{Pastor01a}
 R. Pastor-Satorras and A. Vespignani, Phys. Rev. Lett. {\bf 86}, 3200 (2001)
 
 \bibitem{Pastor01b}
 R. Pastor-Satorras, A. Vespignani, Phys. Rev. E {\bf 63},
066117 (2001)



 \bibitem{Pastor02}
 R. Pastor-Satorras and A. Vespignani, Phys. Rev. E {\bf 65}, 035108R (2002)

 \bibitem{Pastor15}
R. Pastor-Satorras,  C. Castellano, P. Van Miegh, and A. Vespignani, Rev. Mod. Phys. {\bf 87}, 925 (2015) 

 \bibitem{Marro99}
 J. Marro and R. Dickman, {\it Nonequilibrium Phase Transitions
in Lattice Models} (Cambridge University Press, Cambridge,
1999)
 
\bibitem{Boguna02}
M. Bogu\~n\`a  and R. Pastor-Satorras, Phys. Rev. E {\bf 66}, 047104 (2002)

\bibitem{Morreno02}
Y. Moreno, R. Pastor-Satorras, and A. Vespignani, Eur. Phys.
J. B {\bf 26}, 521 (2002)

\bibitem{Morreno03}
Y. Moreno and A. V\'azquez, Eur. Phys. J. B {\bf 31}, 265 (2003)

\bibitem{Bart05}
M. Barth\`elemy,  A. Barrat, R. Pastor-Satorras, and A. Vespignani,
J. Theor. Biol. {\bf 235}, 275 (2005)

\bibitem{Mieghem13}
P. Van Mieghem and R. van de Bovenkamp, Phys. Rev. Lett.
{\bf 110}, 108701 (2013)


\bibitem{Chatterjee09}
S. Chatterjee and R. Durrett, Ann. Probab. {\bf 37}, 2332 (2009)

\bibitem{Ferreira12}
S. C. Ferreira,  C. Castellano, and R. Pastor-Satorras, Phys. Rev. E {\bf 86}, 041125 (2012)

\bibitem{Mata13}
A. S. Mata and S. C. Ferreira,  Europhys. Lett. {\bf 103}, 48003 (2013)

\bibitem{Fine11}
P. Fine, K. Eames, and D.L. Heymann,  Clin. Infect. Dis. {\bf 52},   911 (2011) 

\bibitem{Pastor02b}
R. Pastor-Satorras and A. Vespignani,  Phys. Rev. E {\bf 65},
036104 (2002)

\bibitem{Cohen01}
R. Cohen,  K. Erez, D. ben-Avraham, and S. Havlin,  Phys. Rev.
Lett. {\bf 86}, 3682 (2001)

\bibitem{Cohen03}
R. Cohen,  S. Havlin, and D. ben-Avraham, Phys. Rev. Lett.
{\bf 91}, 247901 (2003)

\bibitem{Holme02}
P. Holme,  B. J. Kim, C. N. Yoon, and S. K. Han, Phys. Rev. E
{\bf 65},  056109 (2002)



\bibitem{Holme04}
P. Holme,  Europhys. Lett. {\bf 68}, 908 (2004)

\bibitem{Bender78}
E. A. Bender and E. R. Canfield,  J. Comb. Theory Ser. A {\bf 24},
296 (1978)

\bibitem{Molloy95}
M. Molloy and B. Reed, Random Struct. Algorithms {\bf 6}, 161 (1995)

\bibitem{Molloy98}
M. Molloy and B. Reed, Combinatorics Probab. Comput. {\bf 7},
295 (1998)

 \bibitem{Cohen00}
R. Cohen,  K. Erez, D. ben-Avraham, and S. Havlin,  Phys. Rev.
Lett. {\bf 85}, 4626 (2000)


\bibitem{Catanzaro05}
M. Catanzaro, M. Bogu\~n\'a, and R. Pastor-Satorras, Phys. Rev.
E {\bf 71}, 027103 (2005)

\bibitem{Hoshen}
J. Hoshen and R. Kopelman, Phys Rev E {\bf 14}, 3438 (1976)

{  

\bibitem{Blavatska08}
V. Blavatska and W. Janke, J. Phys. A {\bf 42}  015001 (2009)

\bibitem{Lapshina19}
S. Yu.  Lapshina,  Lobachevskii J. Math, {\bf  40} 341 (2019)

\bibitem{Kotwica19}
M.  Kotwica, P.  Gronek,  and K.  Malarz, 
Int. J. Mod. Phys.  C {\bf 30 }, 1950055 (2019)

\bibitem{Oliveira20}
F.C.  de Oliveira, S.  Khani, J.M. Maia, and F.W. Tavares, 
Mol. Simul. {\bf 46}, 1453 (2020)
}



\bibitem{Stauffer}
D. Stauffer and A. Aharony, {\it Introduction to Percolation
Theory} ( Taylor and Francis London 1992)



\bibitem{Sykes}
M.F. Sykes and M. Glen, J. Phys. A: Math. Gen. {\bf 9}, 87 (1976)


\bibitem{Berche09}
 B. Berche, C. von Ferber, T. Holovatch, and Yu. Holovatch,  Eur. Phys. J. B {\bf 71},  125 (2009)

\bibitem{Berche12}
 B. Berche, C. von Ferber, T. Holovatch, and Yu. Holovatch,
  	Adv. in Complex Systems {\bf 15},  1250063 (2012)






\bibitem{Milligan15}
 G.N Milligan and  D.T. Barrett {\it Vaccinology: an essential guide} (Chichester, West Sussex: Wiley Blackwel, 2015)

\bibitem{Grassberger83}
P. Grassberger, Math. Biosc. {\bf  63}, 157 (1983)

\bibitem{Gagliardi10}
H F Gagliardi and D. Alves, Math. Popul. Stud. {\bf 17}, 79 (2010)



\end{thebibliography}
\end{document}